\documentclass{elsart}

\usepackage{graphicx}
\usepackage{amssymb}

\newcommand{\eg}{{\it e.g.\,}}
\newcommand{\ie}{{\it i.e.\,}}

\newcommand{\f}{\frac}
\newcommand{\outness}{D}

\newcommand{\bea}{\begin{eqnarray}}
\newcommand{\eea}{\end{eqnarray}}
\newcommand{\be}{\begin{equation}}
\newcommand{\ee}{\end{equation}}

\begin{document}

\begin{frontmatter}


\title{Centrality properties of directed module members in social networks
%\thanksref{label1}
}
%\thanks[label1]{}
\author[label3]{P{\'e}ter Pollner
%\thanksref{label2}
}
%\thanks[label2]{}
\corauth[cor1]{Corresponding author}


\author[label3]{Gergely Palla}

\author[label4]{D\'aniel \'Abel}

\author[label5]{Andr\'as Vicsek}

\author[label3]{Ill{\'e}s J. Farkas}

\author[label4]{Imre Der{\'e}nyi}

\author[label3,label4]{Tam\'as Vicsek\corauthref{cor1}}
\ead{vicsek@angel.elte.hu}
\author{}
\ead[url]{http://www.cfinder.org}

\address[label3]{Statistical and Biological Physics Research Group of Hungarian 
Academy of Sciences\\
1117 Budapest, P\'azm\'any P. stny. 1/A
}

\address[label4]{Department of Biological Physics, E{\"o}tv{\"o}s University\\
1117 Budapest, P\'azm\'any P. stny. 1/A
}

\address[label5]{Outline Foundation \\
1023 Budapest, Frankel Le{\'o} u. 21-23. II/7
}

\begin{abstract}
% Text of abstract
Several recent studies of complex networks have suggested algorithms for
locating network communities, also called modules or clusters, which
are mostly defined as groups of nodes with dense internal connections.
Along with the rapid development of these clustering techniques, the 
ability of revealing overlaps between communities has become very important 
as well. 
An efficient search technique for locating overlapping modules is provided
by the Clique Percolation Method (CPM) and its extension
to directed graphs, the CPMd algorithm.
Here we investigate the centrality properties of directed module members
in social networks obtained from e-mail exchanges and from 
sociometric questionnaires. 
Our results indicate that nodes in the overlaps between
modules play a central role in the studied systems. Furthermore,
the two different types of networks show interesting differences
in the relation between the centrality measures and the 
role of the nodes in the directed modules.
\end{abstract}

\begin{keyword}
network modules \sep
          overlapping modules \sep
          directed networks \sep
          clustering \sep
          community finding

% http://www.aip.org/pacs/pacs08/pacs0880.html
\PACS 
89.75.Hc % Networks and genealogical trees
\sep 
89.75.Fb % Structures and organisation in complex systems
\sep 
89.65.Ef % Social organisations; anthropology
\sep
89.75.Da % Systems obeying scaling laws
\end{keyword}

\end{frontmatter}

% main text
\section{Introduction}
\label{sec:intro}

% \appendix

% \section{}
% \label{}

A widespread approach to the analysis of complex social and economic
phenomena is to assemble the participating individuals or objects 
and their interactions into a network (nodes and links) and to
infer functional characteristics of the entire system from
this static web of connections
\cite{b-a-revmod,dorog-mendes-book,latora06review,newmanReview,wattsReview}.
Over the past few years, several broadly studied
{\it large-scale} properties of real-world webs
have been uncovered, \eg 
the broad (scale-free) distribution of node degree
\cite{BA99}, overrepresented small subgraphs
\cite{Milo02,Alon07NRG}
and various signatures of hierarchical/modular organisation
\cite{Ravasz02}. Also, many useful measures have been defined 
to quantify the importance of 
the {\it individual nodes} in the networks. 
If a vertex lies on many shortest paths running between other
vertices, it plays a central role in information
flows \cite{brandesbetw} and is responsible for the vulnerability of the
system \cite{netattack}. In a social network the actors capable of 
reaching the others with the lowest number of steps 
(being the closest to others on average) have the
greatest influence \cite{newmanbetw,closenessdef}.

In recent years there has been a quickly growing interest in the 
study of the {\it intermediate-scale} network structures as well.
These units, made up of vertices more densely
connected to each other than to the rest of the network, are often
referred to as communities, modules, clusters or cohesive groups
 \cite{scott-book,pnas-suppl,gn-pnas,palla05nature}.
In the various types of networks' these groups can represent,
communities of people \cite{scott-book,our_new_nature},
functional units in molecular biology \cite{SpirinMirnyPNAS03,ihmels02ng},
a set of tightly coupled stocks or industrial sectors in economy
 \cite{onnela-taxonomy} etc. 
Search techniques allowing overlaps between such modules are becoming
very important \cite{Huberman_PNAS,Baumes_atfed_coms,Guldener,
Vicsek_atfed,Rios_atfed,Zhang_com_overlap_PRE,Newman_atfed_coms,
Reichard-Bornhold_PRL,Nepusz_fuzzy_coms}.
Indeed,
communities in real-world graphs are often inherently overlapping:
 each person in a social web belongs usually  to several groups
(family, colleagues, friends, etc.) and proteins in a protein
interaction 
network
may participate in multiple complexes.
Modules, and also some small subgraphs,
are appropriate for ``coarse-graining'' complex networks:
each module/subgraph can be represented as a node and two such nodes can
be linked,
if the corresponding modules/subgraphs are connected (or overlap)
 \cite{palla05nature,makse05nature,pollner}.

In this paper we study the centrality properties of module members
(nodes) in
 four social networks representing either e-mail connections between
 people or social links obtained from questionnaires. Both types of data sets
 can be represented by {\em directed networks}, therefore, the modules
 are located using a directed community finding method, the
 Directed Clique Percolation Method (CPMd) \cite{dircpm}. Our focus is on
 the correlations between general single-node properties (\eg the closeness)
and the role/importance of the members in the communities, 
(\eg the number of communities they participate in).

\section{Directed network modules}
\subsection{Directed clique percolation method (CPMd)}

\begin{figure}
\centering
{\includegraphics[angle=0,width=0.69\columnwidth]{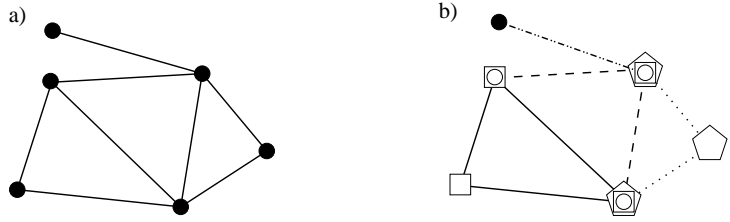}} %.eps
\caption[]{Visualisation of the CPM method. A community is explored
by walking over neighbouring building blocks. The blocks are triangles
in this example, that are $k$-cliques with size $k=3$. At figure a)
the graph is depicted without marking of blocks. At figure b) the
same graph is plotted with marking the nodes with different symbols
for each different block. This graph contains three blocks, and one
node, which is not a member of any blocks. The nodes of each blocks
are marked by circles, rectangles and pentagons respectively. The
unclassifiable node is marked by a filled ellipse. The block built from
rectangles and the block of circles are neighbours since they differ
only in one node. The block of pentagons and circles are neighbours as
well. This graph contains only one community, since all blocks are
reachable from each other by walking over neighbours. In this case the
walks take at most two steps \eg from block of rectangles to block of
circles and than to block of pentagons.}
\label{fig:templroll}
\end{figure}

The CPMd is a natural extension of the Clique Percolation Method (CPM)
 \cite{
palla05nature%
,DerenyiPRL%
}. 
In the CPM a community is built up from adjacent blocks of the same
size $k$. These blocks correspond to
$k$-cliques, that are subgraphs with the highest possible density: each 
pair selected from the $k$ nodes of the subgraph is connected.
Two blocks are considered adjacent if they overlap with each other 
as strongly as possible, \ie, they share $k-1$ nodes. A community
is a set of blocks that can be reached from one to the other through a
sequence of adjacent blocks\cite{DerenyiPRL}. 
Note, that any block belongs always to
exactly one community, however, there can be nodes belonging to several
communities. (E.g.\ if blocks overlap only in a single node.) 
The CPM community searching method is visualised in
Fig.\ref{fig:templroll}: the communities found by CPM contain only
nodes that are densely connected. Nodes with only a few connections or
nodes that do not participate in a densely connected subgraph are not
classified into any community.

The CPM method is robust against of removal or insertion of a single
link. Due to the local nature of this approach, such 
perturbations can alter only the communities containing at 
least one of the end points of the link. (In contrast, for global methods
like modularity optimisation the removal or insertion of a single link
can result in the change of the overall community structure.) 

Finally, we note that the CPM will find the same communities in a
given subgraph irrespective to the fact whether the subgraph 
is linked to a larger network or not. Therefore, a heterogeneous
network can be analysed by first dividing it into homogeneous
 parts, and applying the method to these subnetworks separately. 
Homogeneous parts of a network  can be 
determined e.g.\  from external informations 
like geographical locations of the nodes etc.

\begin{figure}
\centering
{\includegraphics[angle=0,width=0.69\columnwidth]{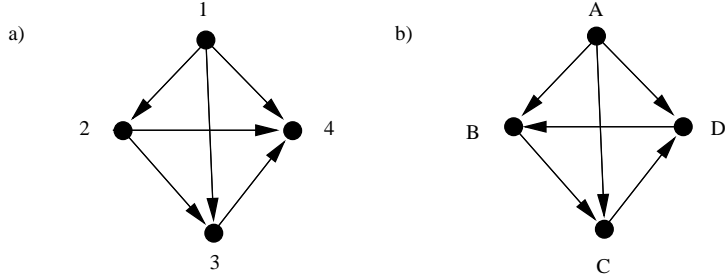}} %.eps
\caption[]{Two example for directed graphs, that are $k$-cliques of
  size $k=4$ if neglecting the directionality of the links. However,
  the graph a) forms a directed $k$-clique, but the graph b) does
  not. The numbers near the nodes of graph a) show the ordering among
the nodes for defining the overall directionality of the clique. The
graph b) cannot be ordered due to the loop formed by the links among
nodes $B$, $C$ and $D$.}
\label{fig:dirclique}
\end{figure}

\begin{figure}
\centering
{\includegraphics[angle=0,width=0.69\columnwidth]{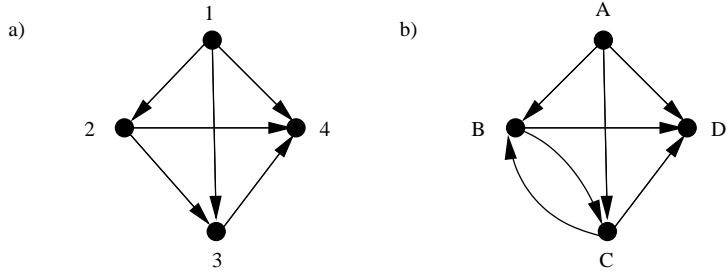}} %.eps
\caption[]{Two example for directed graphs, that are $k$-cliques of
  size $k=4$ if neglecting the directionality of the links. The graph
  a) qualifies as a directed $k$-clique, but the graph b) without
  rectification does not. The double link between the nodes $B$ and
  $C$ forms a loop.} 
\label{fig:dircldouble}
\end{figure}

The undirected approach detailed above can be made inherently directed
by replacing the (undirected) building blocks with 
{\em directed $k$-cliques}. In this
\emph{C}lique \emph{P}ercolation \emph{M}ethod with \emph{d}irected
cliques (CPMd) the building blocks of a community are defined as
complete subgraphs of size $k$ in which an ordering can be made such
that between any pair of nodes there is a directed
link pointing from the node with the higher order towards the lower one
 \cite{dircpm}. See Fig.\ref{fig:dirclique}.
In the absence of bidirectional
 links (also called double links) the above condition is equivalent
 to 
\begin{itemize}
\item the absence of directed loops (closed directed paths) in the directed $k$-cliques.
\item the directionality of all links within the $k$-clique, 
where any directed link points from a node with a higher order to a lower one.
\item the topology, where the number of the nearest out-neighbours
  (neighbours reached along an out-link) is different for each node in
  the $k$-clique.
\end{itemize}
In the presence of bidirectional links a complete subgraph of size $k$ is
qualified as a directed $k$-clique if it is possible to rectify the
bidirectional links in such a way that the remaining unidirectional
links fulfil the above criterion. Note, that without rectification a
$k$-clique with a double link cannot satisfy the definition of
the directed $k$-clique. One can easily see on Fig.\ref{fig:dircldouble},
that above conditions cannot fulfilled \eg there is a loop among
the nodes.

The overall directionality of such an object
naturally follows the ordering of the nodes:
the node with highest order is the one which has only
out-neighbours, and can be viewed as the
``source'' or ``top''-node of the $k$-clique, whereas the node with
lowest order has only incoming
 links from the others, and corresponds to a ``drain'' or ``bottom'' node.

The communities are built up similarly  to the undirected case:
two directed building blocks ($k$-cliques) are considered to be
neighbours if they share $k-1$ nodes. A
directed $k$-clique module (the CPMd modules) arises as the union
of directed $k$-cliques that can be reached from each other through
 a series of neighbouring $k$-cliques.

There are many other possibilities to define a directed $k$-clique. We
have used one particular choice\cite{dircpm}, which is more
restrictive as other natural possibilities, and which is suitable for
any $k$-clique without need of further considerations. Previous results
showed that even this restrictive definition can 
result in similar community structures as the undirected
method\cite{dircpm}. 

\subsection{Quantifying the ranking of directed module members}

A natural and simple approach to rank nodes in a module is to 
compare the number of their incoming and outgoing links. 
For example, a node having only out-neighbours
can be viewed as a ``source'', whereas a node with only incoming links 
as a ``drain''. Most nodes, however, fall between these
two extremes. To quantify this property, we use the 
{\it relative out-degree} of node $i$ in module $\alpha$ as
\be
\outness_{i,{\rm out}}^{\alpha}= \f{d_{i,{\rm out}}^{\alpha}}{d_{i,{\rm in
}}^{\alpha}+
d_{i,{\rm out}}^{\alpha}},
\label{eq:outness}
\ee
where $d_{i,{\rm in}}^{\alpha}$ and $d_{i,{\rm out}}^{\alpha}$ denote 
the numbers
of  in-neighbours and out-neighbours amongst the other nodes in the module,
respectively \cite{dircpm}. 

Obviously, the values of $\outness_{i,{\rm out}}^{\alpha}$ 
fall into the range between 0 (for drain nodes) and 1 (for source
nodes). The analogous quantity for in-degrees is obtained as
$1-D_{i,{\rm out}}^{\alpha}$. 
Since modules may overlap, nodes may have as many relative out(in)-degree as
many modules they participate in.

\section{Centrality measures of individual nodes}

In the following we shall use the {\it betweenness centrality} 
\cite{%
betwdef%
} and the {\it closeness centrality} \cite{closenessdef} to
characterise the role and position of the nodes in the studied networks.
We apply these measures to directed networks, where moving along an edge
is allowed only into the direction of the edge. This restriction effects
the possible shortest paths and the distance of nodes as well.

The betweenness centrality, $b_i$, of node $i$ measures the importance 
of $i$ in the communication between any node pair in the
network \cite{newmanbetw}. We use the definition 
\be
b_i=\sum_{l\ne i\ne j}{\sigma_{lj}(i)},
\ee
where $\sigma_{lj}(i)$ is the portion of number of all shortest paths 
between node $l$ and node $j$ that go through $i$. 
The closeness centrality, $c_i$, of node $i$ is defined as
\be
c_i=\frac{1}{\sum_{j}d(i,j)},
\ee
where $d(i,j)$ is the distance between $i$ and $j$. The
 closeness centrality of a node is large if it can be
 reached within a small number of steps on average.
 Note that in directed networks
the distance (the number of steps along a shortest path)
may depend on the direction of the path.
Therefore, we shall differentiate between in- and out-closeness 
($c_{i,{\rm in}}$ and $c_{i,{\rm out}}$) depending on whether 
$i$ is the end point or the starting point of the path connecting $i$ and $j$.

\section{Results in social networks}

In this section we examine the centrality properties of nodes
in directed social networks with respect to their membership number $m$
(the number of modules they participate in) and with respect to
their relative out-degrees. The networks under study 
represent the directed e-mail connections between the students at the 
University of Kiel \cite{email}, (containing 
$57,158$ nodes and $103,701$ links), the e-mails between the employees of
the Enron concern \cite{enron}, (consisting of
$151$ nodes and $1364$ links), and two networks obtained 
from sociometric questionnaires in a smaller and a medium-sized
company. In the latter cases a directed link was 
set between two employees if one has named the other in any 
question\footnote{In a questionnaire there are several questions concerning 
different type of social relations, and answers are typically names 
of colleagues who are thought to satisfy the posed situation.}, 
resulting in 1166 links between 115 employees for the small company 
and 4349 links between 597 nodes for the other company. 

When extracting the directed modules using the CPMd, we adjusted the
parameter $k$ ($k$-clique size) using a criterion to 
find a modular structure as highly structured as possible. This
criterion is based on the observation, that for many real networks there
is a well defined parameter range, where no giant community (having
size in order of magnitude as the size of the whole 
network) can be found, instead many smaller communities occur.
However, if the parameters are changed (in our case the value of $k$ is
decreased), the rich community structure disappears
and one giant community dominates the network. In the optimal setting 
the parameters are adjusted just below the critical point of this
percolation transition (for details see \cite{palla05nature}).

For data sets studied in this paper the percolation threshold can be
determined by observing the change in the size, $s$, of the largest
community as the control parameter (the clique size, $k$) increases. Since
there is no weight defined on the links in our networks, we have only
one control parameter. Below the
percolation threshold (at small $k$ values) a giant community exists
in the network. If $k$ increases, near the critical point the size
of the largest community drops quickly. At large $k$ values the
size of the largest community decreases at a moderate rate.

\begin{figure}
\centering
{\includegraphics[angle=0,width=0.69\columnwidth]{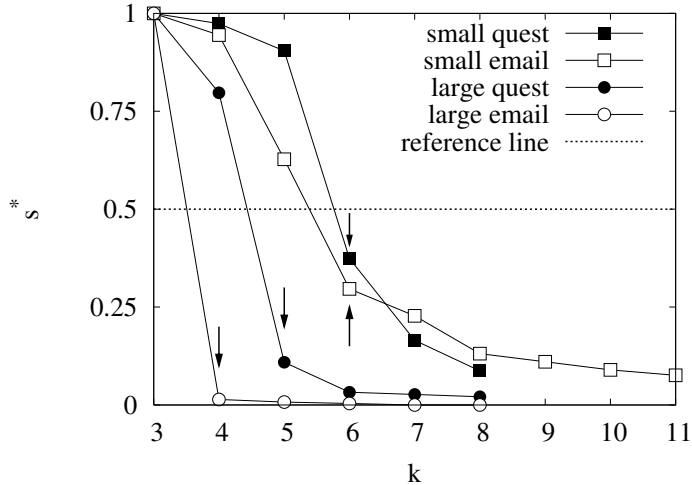}} %.eps
\caption[]{The rescaled size, $s^*$, of the largest community 
decreases with increasing $k$. Near the percolation threshold
$s^*$ decreases rapidly, and the optimal $k$ value (marked by a small
arrow) is chosen to be the
smallest $k$ where no giant community can be found in the network. 

}
\label{fig:optk}
\end{figure}

In Fig.\ref{fig:optk} we show the behaviour of the size of the largest
community. Since the size of the largest community,
$s$, is different for each network, we have plotted a rescaled
community size, $s^*$, equal to $s$ divided by the size of the
largest community at $k=3$. (This way, each curve starts with $s^*=1$ 
 at the smallest possible $k$, which is $k=3$.)
We found, that the optimal $k$ value for the Enron data set
and for the questionnaire data set of the smaller company is
$k=6$. For the larger company we found the optimal value 
$k=5$, whereas for the email data set of the University of Kiel it
turned out to be $k=4$.

\subsection{Extracting networks from email and questionnaire data}

There is a growing interest of measuring and analysing social relation
data within organisations. Relations between people can be of various
types: they can exchange informations, meet each other by lunch,
etc. 

Nowadays information is mainly exchanged by using some electronic
form: either by phone calls\cite{our_new_nature} or by
emails. Email is the electronic form of  letters: there is
always a sender, whom the email is sent from, and there is a receiver,
whom the email is sent to. If one person writes an email to another one, a
social relation emerges between two actors. These contacts can be
represented by a network in which the nodes correspond to the users
and the links to the emails. These connections are directed, and the
natural choice for the direction is to follow the path of the email,
pointing from the sender to the receiver.

The same message content can be spread
among several recipients by listing more than one recipient in the
header. We consider this type of contacts as several individual
relations originating from the sender to each of the receivers. Here
we do not analyse the correlations emerging from emails with multiple
recipients. 

Other, less formal relations (\eg someone can trust in the opinion of
others) are hard to measure directly and are practically impossible to
extract from automatically recorded resources. Thus, one has to apply
indirect measuring methods. The most often used solution is provided
by questionnaires.
In questionnaires the investigated set of people (\eg employees of a
company) are asked the same questions relating to their social
contacts. These questions are formulated along the rules of standard 
sociometry: they must take into account, that the answerers are human
beings with restricted time and memory capacity as compared to
computers. Humans can have different interests and can cheat in
their answers. Well organised questionnaires can help to minimise false
results due to such effects.  

We used results of two surveys, where the employees received a set of
questions concerning their social contacts. The answers to the
questions contained a list of names, where the names had to be picked
out from a prepared list of colleagues.
The network representation was built up from the questionnaire results
by representing the employees as nodes. Two nodes are connected, if
there exists an answer in any of the questionnaires, that 
justifies the existence of a social relation between the corresponding
two employees. If an employee marked another employee in any of the
answers then a directed link points from the answerer to the colleague
named in the answer. 

\subsection{Nodes in community overlaps are central}

After extracting the directed modules using the CPMd, we first
examined the centrality properties of the nodes in the
overlaps between the modules (the nodes having a membership
larger than one). 

\begin{figure}
\centering
{\includegraphics[angle=0,width=0.49\columnwidth]{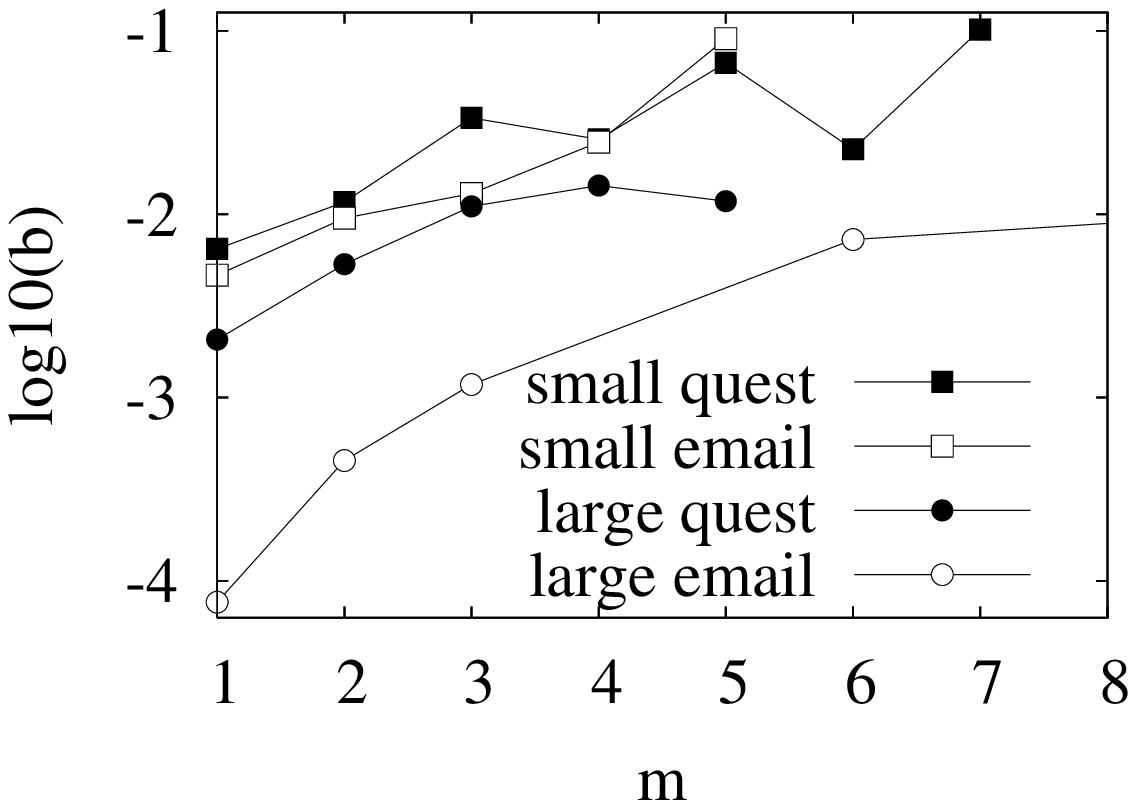}} %.eps
{\includegraphics[angle=0,width=0.49\columnwidth]{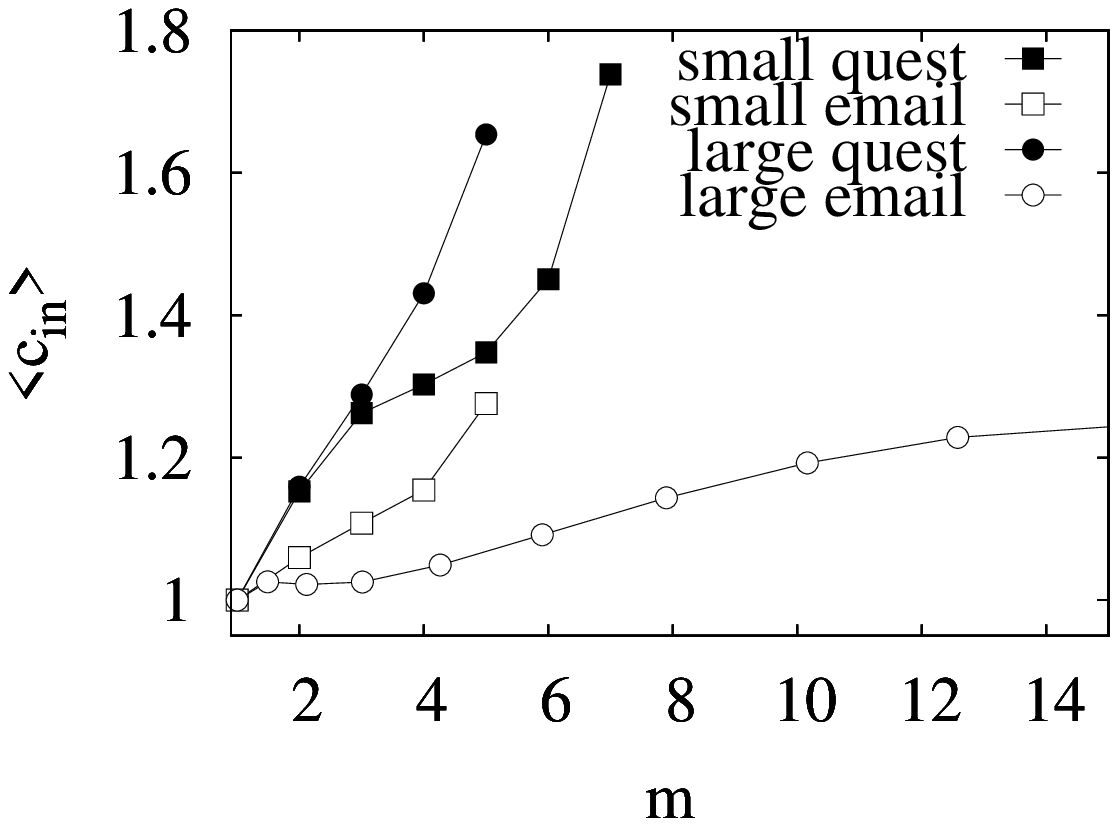}} %.eps
\caption[]{Left: betweenness as a function of the membership number. 
Due to the logarithmic scale nodes with zero betweenness are dropped. The 
smooth lines show the tendency of the data. 
Right: smoothed average in-closeness against the membership number. 
(For clarity the lines are scaled to start at 1).
Both centrality measures show increasing tendency in function of $m$.
}
\label{fig:mem_centr}
\end{figure}

In Fig.~\ref{fig:mem_centr} we show $\left<b_i\right>$ and
 $\left<c_{i,{\rm in}}\right>$ as a function of the membership number $m$. 
We can observe a clear tendency: nodes participating in more communities are 
more central. They can control more effectively the flow of information, 
since they lie typically on
more shortest paths. Also, they can reach other nodes faster, since
they need less steps on average to others.

\begin{figure}
\centering
{\includegraphics[angle=0,width=1.\columnwidth]{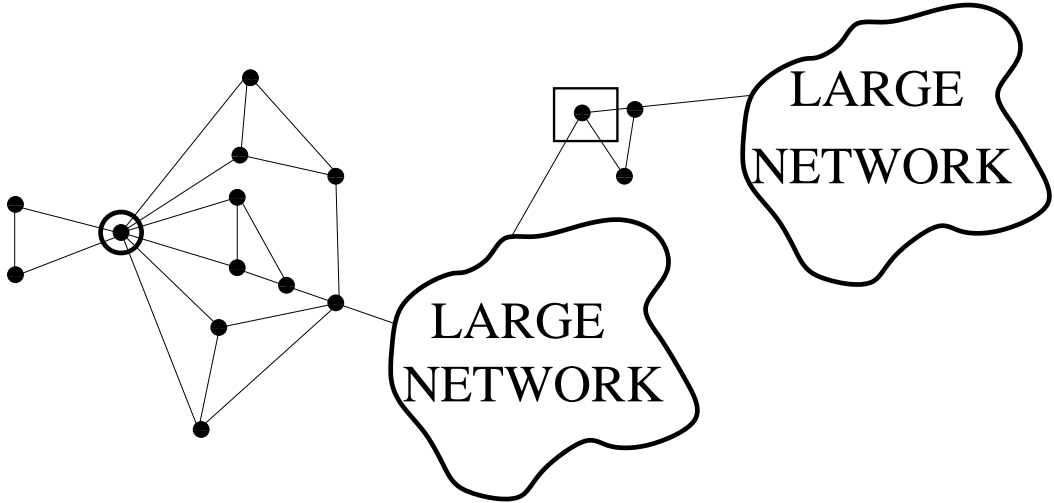}} %.eps
\caption[]{
Counter-example for the correlation between the membership and 
centrality measures. The node marked by a surrounding circle has a
large membership number: at $k=3$ its membership number is $m=5$ and the
betweenness and closeness value of this node is small, since this node is 
at the periphery of the network. However, the node marked by a frame
has a membership number $m=1$ and its betweenness is large, since it  
acts as a bridge.
}
\label{fig:mem_centr_cex}
\end{figure}

Note, that the observed positive correlation between the centrality measures 
and the membership number is not trivial. It is easy to 
construct counter-examples, where nodes with high membership number are at the
periphery of the network. We show a sketch of such a network in 
Fig.~\ref{fig:mem_centr_cex}.
Similar counter-examples can be generated for any $k$.

\subsection{Centrality as a function of the relative out-degree}

\begin{figure}
\centering
{\includegraphics[angle=0,width=0.49\columnwidth]{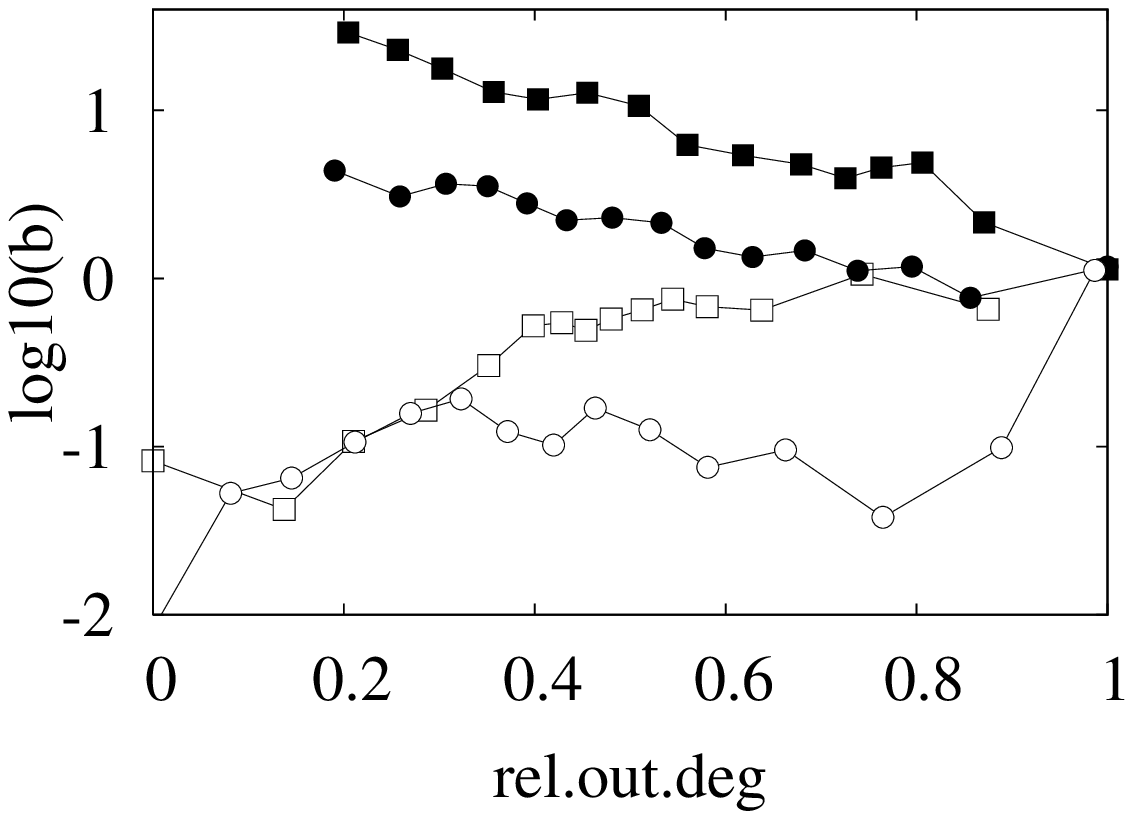}} %.eps
{\includegraphics[angle=0,width=0.49\columnwidth]{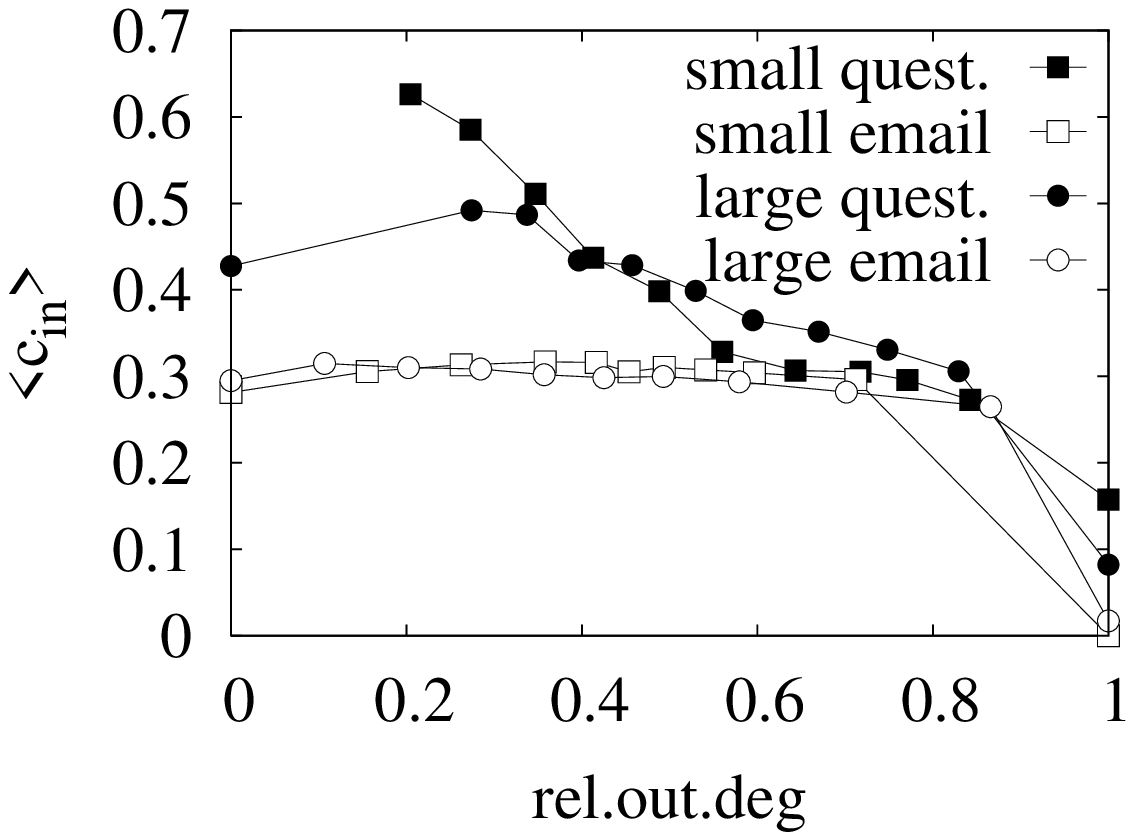}} %.eps
\caption[]{
Left: relative out-degree and betweenness. The smoothed curves are
shifted to 0 at the maximal values. Again, e-mail networks do not show
decreasing tendency as questionnaire networks do.
Right: relative out-degree and averaged in-closeness.
The $\left<c_{i,\rm in}\right>$ is decreasing with $D_{i,{\rm out}}^{\alpha}$
 for the questionnaire networks, whereas it is more or less constant
 for the e-mail networks. 
For better visibility the smoothed in-closeness values are
scaled to take similar values before the cutoff.
}
\label{fig:rout_bet_clin}
\end{figure}

Next we analysed the dependence of the two chosen centrality measures on
 the relative out-degree of the module members.
According to Fig.~\ref{fig:rout_bet_clin} both the betweenness centrality
and the in-closeness show a decreasing tendency with increasing 
$D_{i,{\rm out}}^{\alpha}$ in the networks obtained from
questionnaires (black symbols). In other words:
people named by fewer colleagues than vice versa
are less central in the network.
These people can be reached by others in more steps on average 
(they have smaller in-closeness values), and they have less
control on the information flow (their betweenness centrality is
lower). Interestingly this tendency cannot be observed in 
case of the e-mail networks (white symbols in Fig.~\ref{fig:rout_bet_clin}),
where the centrality measures are more or less independent
of the relative out-degree (or show a slight increasing tendency). For
all four networks $\left<c_{i,\rm in}\right>$ 
suddenly drops to zero at $D_{i,\rm out}^{\alpha}=1$. This is due to the fact
that nodes having only out-neighbours have a zero in-closeness by definition.

This difference between the two types of networks is likely to originate
in the different methods used to 
collect the data.
E-mail networks are generated automatically, no interaction
with the observed persons is necessary. 
In contrast, a questionnaire can not be filled without 
some interaction. Interaction needs
time from the respondent and from the interviewer as
well. This sets some restrictions to the number of emerging links: the
number of questions and the number of possible answers can not be as
large as the number of e-mails between actors that can be exchanged during 
several months. 
Furthermore, e-mails are collected in a wide time span 
(during which fluctuations are averaged out),
whereas a questionnaire
 captures an instantaneous state.

\subsection{Average membership as a function of the relative out-degree}

\begin{figure}
\centering
{\includegraphics[angle=0,width=0.68\columnwidth]{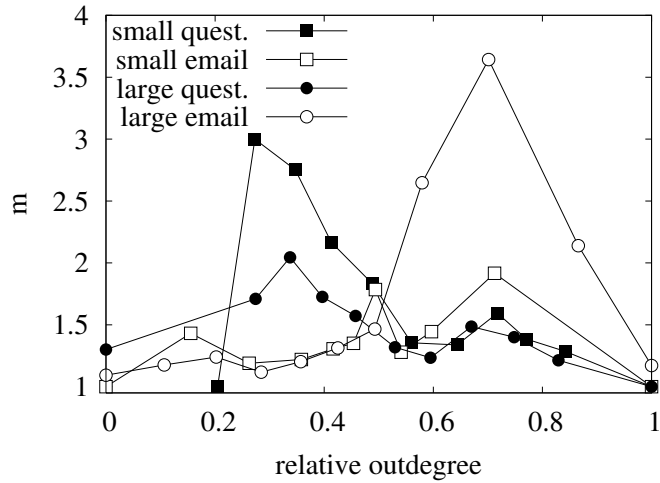}} %.eps
\caption[]{
Relative out-degree and membership: the averaged membership takes its
maximum at lower relative out-degree in questionnaire networks and it is
maximal at high relative out-degree for email networks.
}
\label{fig:mem_rout}
\end{figure}

Another interesting difference between the studied email and
questionnaire networks can be observed in the relation between the average 
membership of a node (the average
number of communities the node participates in) and 
the relative out-degree (\ref{eq:outness}). 
In questionnaire networks the average membership takes larger values
at nodes whose relative out-degree is small. In contrast,
we find the membership maximum at large relative out-degrees for nodes
in email networks. The corresponding curves 
are shown in Fig.\ref{fig:mem_rout}. 

A plausible explanation of this observation
can be that in email networks the directionality of the links follows
the direction of the information flow: there is always some content
in the text body of the email. Those persons, that participate in
several groups (communities) will probably more often forward
information between communities: if some knowledge is available only
in one community, than the actors in the community overlaps will provide 
this information for the other communities they participate in.

In questionnaire networks the
direction is opposite: if one has named somebody he will probably ask
him to receive some information. Hence the direction of the links
are the opposite as of the information flow: the most often named
persons are most trusted for helping or providing helpful information.
If a person is named by many others, than its node will have large
relative in-degree (small relative out-degree). Probably those
persons can help others who can easily build new social contacts, and
therefore they probably participate in several different communities.

\section{Summary}

We have examined four directed social networks constructed from
e-mail exchanges and from questionnaire data. 
We have extracted the overlapping directed module structure of these nets 
using the CPMd algorithm, and studied the behaviour of the 
betweenness centrality and closeness for module members.
Our results indicate that nodes with high memberships (participating
 in numerous modules in parallel) play a central role in these networks.
Furthermore, our investigations revealed interesting differences
 between the two types of networks concerning the dependence of the 
centrality measures on the relative out-degree of module members
(the ratio of out-degree versus number of all nearest neighbours within
the modules) .

\section*{Acknowledgements}

The authors thank the partial support of the Hungarian National Science
Fund (OTKA T034995, K068669, PD048422, K060456) and the National Research
and Technological Office (NKTH, CellCom RET, Textrend). 
Questionnaire data were generously made available by Outline Foundation.

\end{document}